\documentclass[12pt]{article}
\usepackage{graphicx}

\def\Re{{\cal R \mskip-4mu \lower.1ex \hbox{\it e}\,}}
\def\Im{{\cal I \mskip-5mu \lower.1ex \hbox{\it m}\,}}
\def\ie{{\it i.e.}}

\def\grtsim{\,\,\rlap{\raise 3pt\hbox{$>$}}{\lower 
3pt\hbox{$\sim$}}\,\,}

\def\etal{{\it et al.}}

\def\sub#1{_{\lower.25ex\hbox{$\scriptstyle#1$}}}
\def\tev{\,{\ifmmode\mathrm {TeV}\else TeV\fi}}
\def\gev{\,{\ifmmode\mathrm {GeV}\else GeV\fi}}
\def\mev{\,{\ifmmode\mathrm {MeV}\else MeV\fi}}
\def\mpl{\ifmmode \overline M_{Pl}\else $\overline M_{Pl}$\fi}
\def\to{\rightarrow}

\def\subw{_{\rm w}}
\def\mh{\ifmmode m\sbl H \else $m\sbl H$\fi}
\def\mch{\ifmmode m_{H^\pm} \else $m_{H^\pm}$\fi}
\def\mt{\ifmmode m_t\else $m_t$\fi}
\def\mc{\ifmmode m_c\else $m_c$\fi}
\def\mz{\ifmmode M_Z\else $M_Z$\fi}
\def\mw{\ifmmode M_W\else $M_W$\fi}
\def\mws{\ifmmode M_W^2 \else $M_W^2$\fi}
\def\mhs{\ifmmode m_H^2 \else $m_H^2$\fi}   
\def\mzs{\ifmmode M_Z^2 \else $M_Z^2$\fi}
\def\mts{\ifmmode m_t^2 \else $m_t^2$\fi}
\def\mcs{\ifmmode m_c^2 \else $m_c^2$\fi}
\def\mchs{\ifmmode m_{H^\pm}^2 \else $m_{H^\pm}^2$\fi}
\def\ztwo{\ifmmode Z_2\else $Z_2$\fi}
\def\zone{\ifmmode Z_1\else $Z_1$\fi}
\def\mtwo{\ifmmode M_2\else $M_2$\fi}
\def\mone{\ifmmode M_1\else $M_1$\fi}
\def\tb{\ifmmode \tan\beta \else $\tan\beta$\fi}
\def\xw{\ifmmode x\subw\else $x\subw$\fi}
\def\ch{\ifmmode H^\pm \else $H^\pm$\fi}
\def\lum{\ifmmode {\cal L}\else ${\cal L}$\fi}
\def\inpb{\,{\ifmmode {\mathrm {pb}}^{-1}\else ${\mathrm 
{pb}}^{-1}$\fi}}
\def\infb{\,{\ifmmode {\mathrm {fb}}^{-1}\else ${\mathrm 
{fb}}^{-1}$\fi}}
\def\epem{\ifmmode e^+e^-\else $e^+e^-$\fi}
\def\ppb{\ifmmode \bar pp\else $\bar pp$\fi}
\def\bsg{\ifmmode B\to X_s\gamma\else $B\to X_s\gamma$\fi}
\def\bsll{\ifmmode B\to X_s\ell^+\ell^-\else $B\to X_s\ell^+\ell^-$\fi}
\def\bstt{\ifmmode B\to X_s\tau^+\tau^-\else $B\to X_s\tau^+\tau^-$\fi}
\def\lamt{\ifmmode \tilde\lambda\else $\tilde\lambda$\fi}
\def\shat{\ifmmode \hat s\else $\hat s$\fi}
\def\that{\ifmmode \hat t\else $\hat t$\fi}
\def\uhat{\ifmmode \hat u\else $\hat u$\fi}

\newskip\zatskip \zatskip=0pt plus0pt minus0pt
\def\matth{\mathsurround=0pt}
\def\lsim{\mathrel{\mathpalette\atversim<}}
\def\gsim{\mathrel{\mathpalette\atversim>}}
\def\atversim#1#2{\lower0.7ex\vbox{\baselineskip\zatskip\lineskip\zatskip
  \lineskiplimit 
0pt\ialign{$\matth#1\hfil##\hfil$\crcr#2\crcr\sim\crcr}}}

\def\grtsim{\,\,\rlap{\raise 3pt\hbox{$>$}}{\lower 
3pt\hbox{$\sim$}}\,\,}
\def\lsim{\,\,\rlap{\raise 3pt\hbox{$<$}}{\lower 3pt\hbox{$\sim$}}\,\,}

\renewcommand{\thefootnote}{\fnsymbol{footnote}}

\begin{document} \begin{titlepage} 
\rightline{\vbox{\halign{&#\hfil\cr
&SLAC-PUB-9614\cr
&December 2002\cr}}}
\begin{center} 
\thispagestyle{empty}
\flushbottom

{\Large\bf Brane-localized Kinetic Terms
 in the Randall-Sundrum Model}
\footnote{Work supported in part by the Department of 
Energy, Contract DE-AC03-76SF00515}
\footnote{e-mails:
$^a$hooman@ias.edu , $^b$hewett@slac.stanford.edu, and
$^c$rizzo@slac.stanford.edu}
\medskip
\end{center}

\centerline{H. Davoudiasl$^{1,a}$, J.L. Hewett$^{2,b}$, and
T.G. Rizzo$^{2,c}$} 
\vspace{8pt} 
\centerline{\it $^1$School of Natural Sciences, Institute for Advanced Study, 
Princeton, NJ 08540} 
\vspace{8pt} 
\centerline{\it $^2$Stanford Linear Accelerator Center, Stanford, CA, 94309}

\vspace*{0.3cm}

\begin{abstract}

We examine the effects of boundary kinetic terms in the
Randall-Sundrum model with gauge fields in the
bulk.  We derive the resulting
gauge Kaluza-Klein (KK) state wavefunctions and
their corresponding masses, as well as the KK gauge field
couplings to boundary fermions, and find that they are
modified in the presence of the boundary terms.  In
particular, for natural choices of the parameters, these fermionic 
couplings can be substantially suppressed compared to those 
in the conventional Randall-Sundrum scenario.
This results in a significant relaxation of the bound on the lightest
gauge KK mass obtained from precision electroweak data; we
demonstrate that this bound can be as low as a few hundred GeV.
Due to the relationship between the lightest gauge KK state and the
electroweak scale in this model, this weakened constraint allows
for the electroweak scale to be near a TeV in
this minimal extension of the Randall-Sundrum model with bulk
gauge fields, as opposed to the conventional scenario.

\end{abstract}

\renewcommand{\thefootnote}{\arabic{footnote}} \end{titlepage}

\section{Introduction}

The Randall-Sundrum (RS) model {\cite {RS1}} offers a new approach to 
the 
hierarchy problem. Within this scenario, the disparity between the 
electroweak 
and Planck scales is generated by the curvature of a 
5-dimensional (5-d) background geometry which is a slice of
anti-de Sitter ($AdS_5$) spacetime. 
This slice is bounded by two 3-branes of equal and opposite tensions 
sitting 
at the fixed points of an $S^1/Z_2$ orbifold which are located at
$\phi=0,\pi$ where $\phi$ is the coordinate of the $5^{th}$ dimension. 
The 5-d warped geometry then 
induces an effective 4-d scale $\Lambda_\pi$ of order a TeV on the 
brane at $\phi=\pi$, the so-called TeV brane. 
$\Lambda_\pi$ is exponentially smaller than the effective scale of 
gravity given by the reduced Planck scale, \mpl, with the suppression 
being determined by 
the product of the 5-d curvature parameter $k$ and 
the separation $r_c$ of the two branes, \ie, $\Lambda_\pi=\mpl 
e^{-kr_c\pi}$.
In this theory, the original parameters 
of the 5-d action are all naturally of the size $\sim \mpl$, while in the 
4-d picture a hierarchy with $\Lambda_\pi/\mpl \lsim 10^{-15}$ appears.  
It has been demonstrated \cite{gw} that this scenario is naturally 
stabilized without the introduction of fine-tuning
for $kr_c\simeq 11$, which is the numerical value 
required to generate the hierarchy.  This model
leads to a distinct set of phenomenological signatures that may 
soon be revealed by experiments at the TeV scale; these  have been 
examined in 
much detail {\cite {phen,gauge1,gauge2}}. 

For model building purposes, it is advantageous to extend the original 
framework of the RS model, where gravity alone propagates in the 
extra dimension, to the case where
at least some of the Standard Model (SM) fields are present in the 
bulk \cite{gw2}.  The simplest such possibility is to have the SM gauge 
fields in the bulk {\cite {gauge1,gauge2}}, while
the other SM fields remain localized on the TeV brane. This scenario 
was examined in detail some time ago \cite{gauge1}
where it was found that the couplings of the Kaluza-Klein (KK) 
excitations of the bulk
gauge fields to matter on the boundary were enhanced by a factor of 
$\sqrt {2\pi kr_c}$ compared to those
of the corresponding zero-mode states.  Since 
$kr_c \simeq 11$ as discussed above, this enhancement is numerically
significant and is 
approximately a factor of $\simeq 8.4$.  Severe constraints are then
imposed on the gauge KK excitations from precision 
electroweak measurements; these imply that the mass of the 
lightest gauge KK state must be heavy with 
$m_1\geq 25-30$ TeV. Due to the relation in this model between the 
masses of the KK states and the scale of physics on the TeV brane,
this bound then correspondingly requires that 
$\Lambda_\pi \geq 100$ TeV. This places such a scenario in a very 
unfavorable light in terms of its original motivation of resolving the hierarchy.

Here, we investigate whether Brane Localized Kinetic Terms (BLKT's) for 
bulk gauge fields can modify these results.
Recently, Carena, Tait and Wagner {\cite {ctw}} examined the 
phenomenological influence of such localized kinetic terms for bulk
gauge fields within the context of flat 
TeV$^{-1}$-sized extra dimensions. These authors showed 
that such terms can lead to significant modifications to the gauge KK 
spectrum, as well as to the KK state self-couplings and 
couplings to fields remaining on orbifold 
fixed points.  Localized kinetic terms are expected to be 
present on rather general grounds in any orbifolded theory, 
as was shown by Georgi, Grant and Hailu {\cite {ggh}}, 
since they are needed 
to provide counterterms for divergences that are generated 
at one loop.  In addition, these boundary terms
can have important implications in a number of different 
situations such as model building {\cite {martin}}, the construction of 
GUTs in higher dimensions with flat {\cite {hall}} or warped \cite{new}
geometries, and, in a different context, the 
quasilocalization of gravity {\cite {grav}}. Up to now the
phenomenological implications of 
BLKT's in the case of warped extra dimensions have not been examined.

We examine the effects of localized kinetic terms in the case 
where the SM gauge fields are present in the RS bulk and the
remaining SM fields are on the TeV-brane. As in the case of
TeV$^{-1}$-sized extra dimensions, we find that these boundary terms 
alter both the KK spectrum as well as their associated 
couplings to the remaining 
wall fields, but in a manner which is quite distinctive due to the
curved geometry in this scenario. We will show that for 
a reasonable range of model parameters these modifications can be quite 
significant and lead to a natural reduction in the strength of the KK 
couplings.  This allows the restrictions on $\Lambda_\pi$ to be reduced 
by more than an order of magnitude 
and it would then no longer be unfavorable to have 
gauge fields be the only SM fields present in the RS bulk. 
The next section contains a detailed 
discussion of the modification of the usual gauge boson analysis within 
the RS framework due to the existence of brane localized kinetic terms.
In Section 3 we discuss the implication of these results for the 
phenomenology 
of this scenario and our conclusions can be found in Section 4.
Special limits of the KK mass equation are discussed in an Appendix.

\section{Formalism}

In this section, we derive the wavefunctions and the 
KK mass eigenvalue equation
for bulk gauge fields with BLKT's in the RS scenario.  We then 
compute the coupling of the gauge field KK modes to 4-d
fermions localized on the TeV brane, where the scale of physics is
of order the electroweak scale.  Our derivation will demonstrate how 
the boundary terms
modify these couplings from their original form and result in
a possible softening of the precision electroweak bounds on the
mass of the lightest KK gauge state.  In the rest of the paper, KK modes 
refer to those from the bulk gauge fields, unless otherwise specified.

We perform our calculations for a $U(1)$ gauge field; the
generalization to the case of non-Abelian fields is straightforward.
Let us start with the 5-d action for the gauge sector
\begin{eqnarray}
S_A = &-& \frac{1}{4}\int d^5 x \, \sqrt {-G} \, \left\{
G^{AM}G^{BN}F_{AB}F_{MN} \right.
\nonumber\\
&+& \left. [c_0 \, \delta(\phi) + c_\pi \, \delta(\phi - \pi)] \,
G^{\alpha \mu} G^{\beta \nu}F_{\alpha \beta}F_{\mu \nu}\right\}\,,
\label{SA}
\end{eqnarray}
where the terms proportional to $c_0$ and $c_\pi$ 
represent the kinetic terms for
the branes located at $\phi=0,\,\pi$, respectively, and
the metric $G_{MN}$ is given by the RS line element
\begin{equation}
ds^2 = e^{- 2 \sigma} \eta_{\mu \nu} d x^\mu d x^\nu -
r_c^2 d \phi^2 \, \, \, \, ; \, \, \, \, \sigma = k r_c |\phi|\,.
\label{ds2}
\end{equation}
Here the dimensionless constants $c_{0,\pi}$ would naturally be 
expected 
to be of order unity. Note that we can neglect terms proportional to 
$\partial_5 A_\mu$ on either brane as demonstrated by {\cite {ctw}}.
We will work in the $A_5 = 0$ gauge so that terms including $A_5$ 
can also be neglected. 
Thus only the usual 4-d kinetic term appears in the brane contributions 
to the action. In our notation, $A, B, \ldots = 0, \ldots, 3,5$ and
$\alpha, \beta, \ldots = 0, \ldots, 3$.  Here, $k$ is the
5-d curvature scale, $r_c$ is the compactification radius, and
$x^5 = r_c \phi$ with $-\pi \leq \phi \leq \pi$.
The field strength is given by
\begin{equation}
F_{MN} = \partial_M A_N - \partial_N A_M\,.
\label{FMN}
\end{equation}

The gauge field is assumed to have the KK expansion
\begin{equation}
A_\mu (x, \phi) = \sum_n A_\mu^{(n)} (x)
\frac{\chi^{(n)}(\phi)}{\sqrt{r_c}}\,.
\label{AKK}
\end{equation}
Substituting the above expansion in the action (\ref{SA}) yields
\begin{eqnarray}
S_A = &-& \frac{1}{4 r_c}\int d^5 x
\sum_{m, n} \, \left\{[1 + c_0 \, \delta(\phi) +
c_\pi \, \delta(\phi - \pi)] \,
F^{\mu \nu(m)}F_{\mu \nu}^{(n)}
\chi^{(m)}\chi^{(n)} \right. \nonumber\\
&-& \left. (2/r_c^2) \,
A^{\mu (m)}A_\mu^{(n)} e^{-2 \sigma}[ (d/d\phi)
\chi^{(m)}][(d/d\phi)\chi^{(n)}]\right\}\,,
\label{SAKK}
\end{eqnarray}
where now all the Lorentz indices are contracted using the
4-d Minkowski metric.  In order to cast this action in the
diagonal form, we demand
\begin{equation}
\int d \phi \, [1 + c_0 \, \delta(\phi) +
c_\pi \, \delta(\phi - \pi)] \,
\chi^{(m)}\chi^{(n)} = Z_n \delta^{m n}
\label{diag1}
\end{equation}
and
\begin{equation}
\frac{1}{r_c^2}\int d \phi \, e^{-2 \sigma} \Big[\frac{d}{d\phi}
\chi^{(m)}\Big]\Big[\frac{d}{d\phi}\chi^{(n)}\Big] =
Z_n m_n^2 \, \delta^{m n}\,.
\label{diag2}
\end{equation}
These equations imply
\begin{equation}
\frac{d}{d\phi} \left(e^{-2 \sigma} \frac{d}{d\phi}
\chi^{(n)}\right) + [1 + c_0 \, \delta(\phi) +
c_\pi \, \delta(\phi - \pi)] \,
r_c^2 \, m_n^2 \, \chi^{(n)} = 0\,.
\label{wave}
\end{equation}
Away from the boundaries at $\phi = 0, \pi$, the solution
to Eq.(\ref{wave}) is given by \cite{gauge1,gauge2}
\begin{equation}
\chi^{(n)}(\phi) = \frac{e^\sigma}{N_n} \, \zeta_1(z_n)\,,
\label{chi}
\end{equation}
where $z_n(\phi) \equiv (m_n/k) e^\sigma$; 
$\zeta_q(z_n)\equiv [J_q(z_n) + \alpha_n \, Y_q(z_n)]$, with
$J_q$ and $Y_q$
denoting the usual Bessel functions of order $q$.  The normalization
$N_n$ is set by the condition
\begin{equation}
\int_{-\pi}^{+\pi} d \phi ~[\chi^{(n)}]^2 = 1\,,
\label{ortho}
\end{equation}
implying that the zero mode is given by $\chi^{(0)} =
1/\sqrt{2 \pi}$.
Integrating Eq.(\ref{wave}) around the fixed points $\phi = 0$
and $\phi = \pm \pi$ yields
\begin{equation}
2 \frac{d}{d\phi}
\chi^{(n)}(0) + c_0 \,
r_c^2 m_n^2 \, \chi^{(n)}(0) = 0
\label{fixed1}
\end{equation}
and
\begin{equation}
2 \frac{d}{d\phi}
\chi^{(n)}(\pi) - e^{2 k r_c \pi} c_\pi
r_c^2 \, m_n^2 \, \chi^{(n)}(\pi) = 0\,,
\label{fixed2}
\end{equation}
respectively.

Let $\varepsilon_n \equiv z_n(0)$.  For light KK modes with
masses $m_n \sim 1$ TeV, which are of phenomenological interest,
$\varepsilon_n \ll 1$; for $k \sim \mpl$ we have
$\varepsilon_n \sim 10^{-15}$.  Substituting for $\chi^{(n)}$
from Eq.(\ref{chi}) in Eq.(\ref{fixed1}), and using the identity
\begin{equation}
x\zeta_0(x) = \zeta_1(x)+x\zeta_1'(x)\,,
\end{equation}
we obtain
\begin{equation}
\alpha_n = - \frac{
J_0(\varepsilon_n) + \delta_0 \varepsilon_n J_1(\varepsilon_n)}
{
Y_0(\varepsilon_n) + \delta_0 \varepsilon_n Y_1(\varepsilon_n)}\,,
\label{alphan}
\end{equation}
where $\delta_0 \equiv c_0 k \, r_c/2$ and a prime $(')$ denotes a
derivative.  
Note that with $c_0$ of order unity, we may expect 
values of $\delta_0$ of order 10 since $kr_c$ takes on the
value $\sim 11$ in order to address the hierarchy.  
Similarly, from Eq.(\ref{fixed2}), we obtain
\begin{equation}
\zeta_0(x_n) - \delta_\pi x_n \zeta_1(x_n) =0\,,
\label{xn2}
\end{equation}
where $\delta_\pi \equiv c_\pi k \, r_c/2$ and
$x_n \equiv z_n(\pi)$.  
The roots $x_n$ of Eq.(\ref{xn2}) yield
the gauge KK mass spectrum, $m_n = x_n k e^{-k r_c \pi}$.  We will 
study
various limits of this root equation in the Appendix.
Following the same arguments as above, we may expect that 
$|\delta_{0, \pi}| \lsim {\cal O}(10)$, assuming no fine-tuning.

\begin{figure}[htbp]
\centerline{
\includegraphics[width=9cm,angle=90]{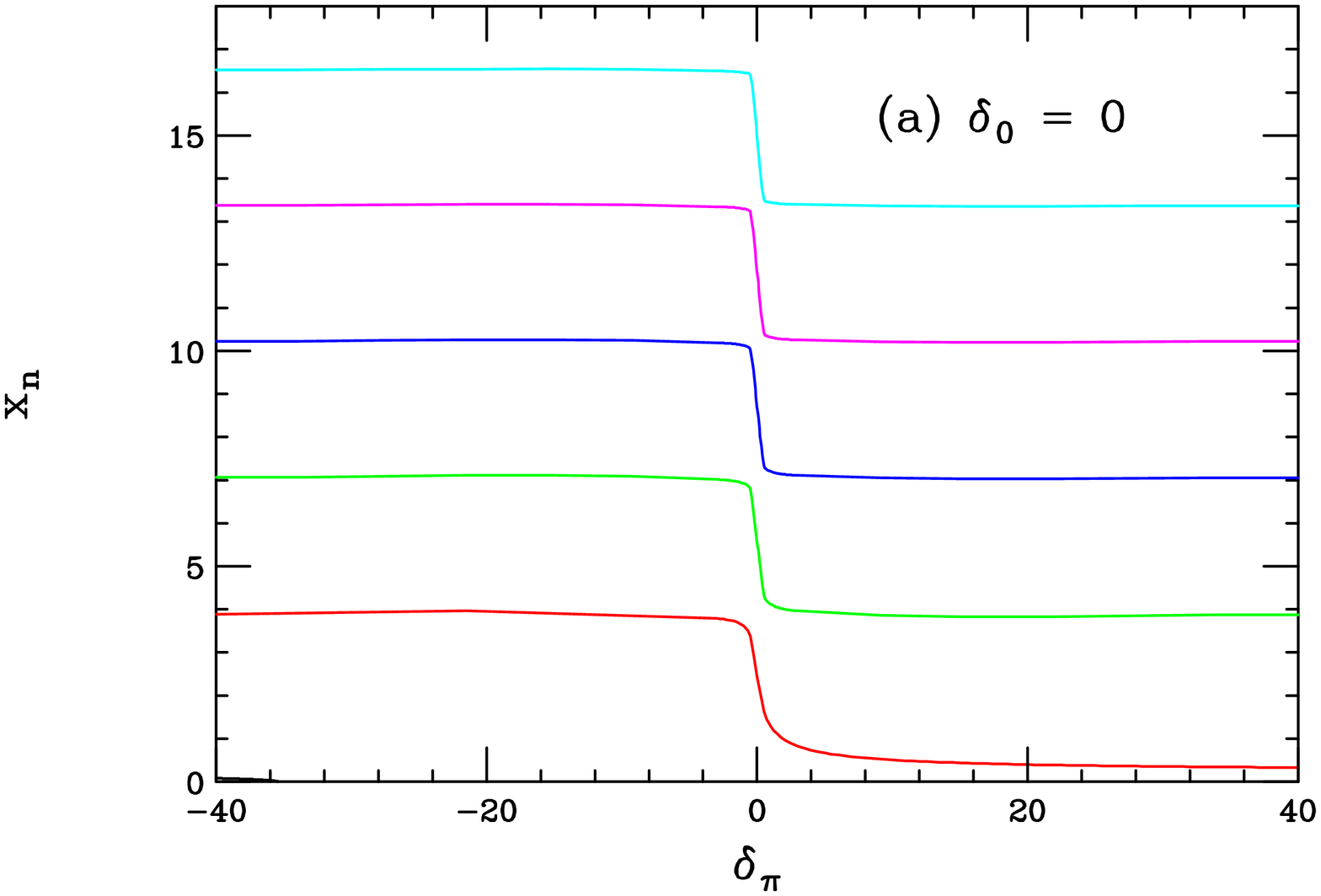}}
\vspace{5mm}
\centerline{
\includegraphics[width=9cm,angle=90]{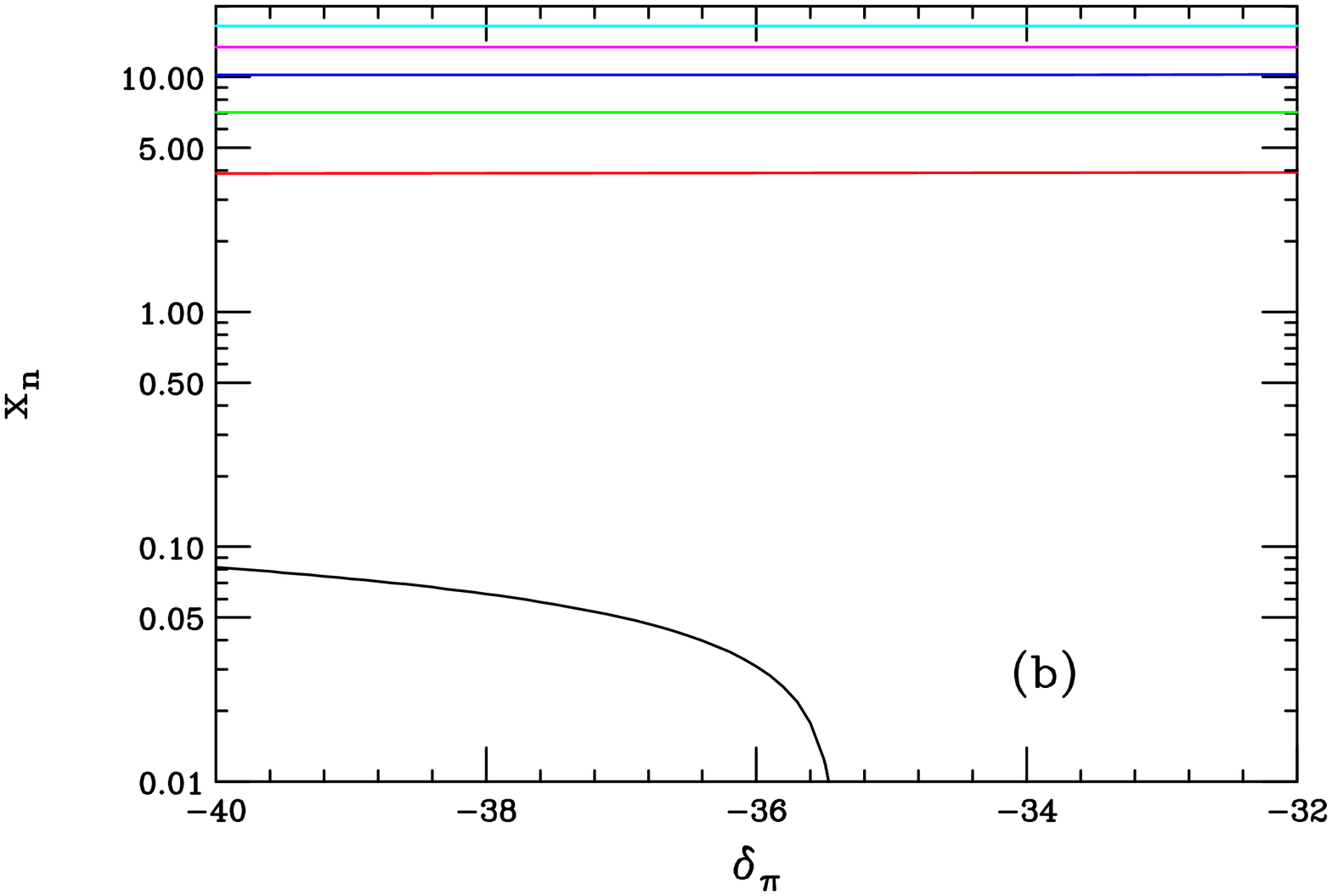}}
\vspace*{0.1cm}
\caption{(a) The behavior of the first five roots as $\delta_\pi$ is 
varied, assuming $\delta_0=0$.  Note the sharp increase in $x_n$
near $\delta_\pi=0$  for all KK
levels.  (b) The large negative $\delta_\pi$
region is expanded to show the new root originating at the value 
$\delta_\pi=
-kr_c\pi\simeq -35.4$.}
\label{fig1}
\end{figure}

Due to the exponentially small value of $\epsilon_n$, it may first
appear that the $\delta_0$-dependent corrections to $\alpha_n$ are 
highly
suppressed, and that $\alpha_n$ is small, of order $\sim 0.01$, as in
the standard RS scenario.  However, an expansion of Eq. (\ref{alphan}) 
in the parameter $\varepsilon_n$ yields
\begin{equation}
\alpha_n \simeq -\frac{\pi/2}
{\ln(x_n/2) - k \, r_c \, \pi + \gamma - \delta_0}\,,
\label{alexp}
\end{equation}
where $\gamma \approx 0.577$ is Euler's constant. $\alpha_n$ remains
small over most regions of parameter space of interest here, except for 
when
$\delta_0\simeq -kr_c\pi\simeq -35.4$.  There are two sources of 
modifications to $\alpha_n$ as compared to the original RS framework
due to the presence of the boundary terms:  (1) the
appearance of the term proportional to $\delta_0$ in the
denominator which is not suppressed by the exponential warp factor, and
(2) the BLKT-modified values of $x_n$. In our numerical analyses we 
will 
use the exact expression for $\alpha_n$, rather than the approximate
version given above, and we take $kr_c=11.27$.

To get a flavor of how the KK tower mass 
spectrum is modified in this scenario with BLKT's, 
we must solve Eq.(\ref{xn2}) for 
a range of values of $\delta_{0,\pi}$.  We first take the
simplest possibility by setting $\delta_0=0$, and examine the roots
$x_n$ as a function of $\delta_\pi$.  
Our results are displayed in Fig. \ref{fig1}.
From Fig. \ref{fig1}a, we see that in the region near $\delta_\pi=0$
the roots undergo a substantial shift but remain relatively
independent of variations in $\delta_\pi$
away from the origin. For large negative values of $\delta_\pi$, the
roots are almost the same as those obtained in the case of the 
graviton KK spectrum in the original RS framework
\cite{phen}. This can be seen analytically by taking the
limit of large $|\delta_\pi|$
in Eq.(\ref{xn2}) and neglecting $\alpha_n\ll 1$.  
When the value of $-\delta_\pi$ exceeds $kr_c\pi$, a new root appears 
which is
barely perceptible in Fig. \ref{fig1}a, but whose turn-on is shown in
detail in Fig. \ref{fig1}b.  The analytical source of this root is
discussed in the Appendix.  We will later see that this signals the
appearance of an unphysical ghost, \ie, a tachyonic state with
negative norm.

\begin{figure}[htbp]
\centerline{
\includegraphics[width=9cm,angle=90]{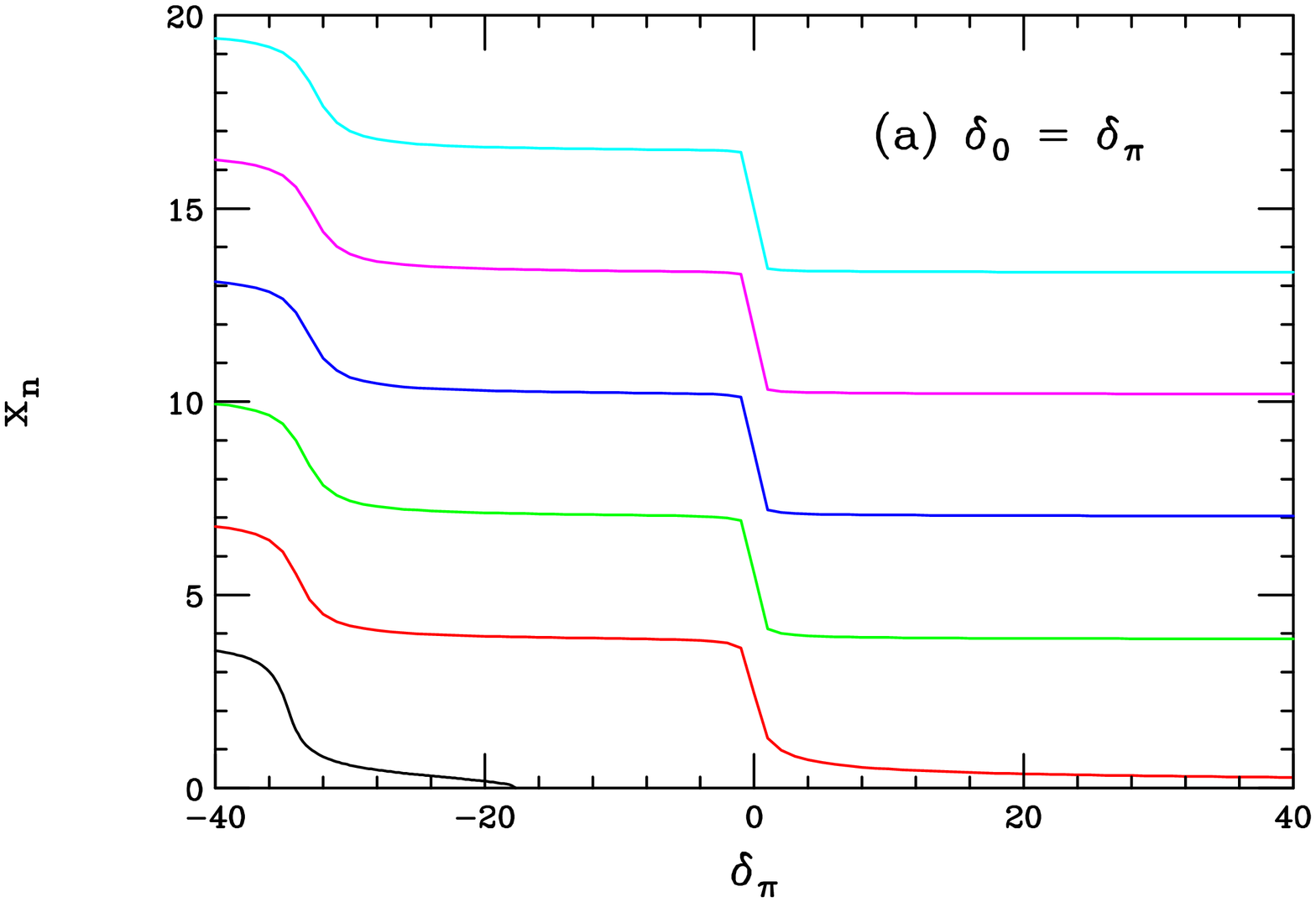}}
\vspace{5mm}
\centerline{
\includegraphics[width=9cm,angle=90]{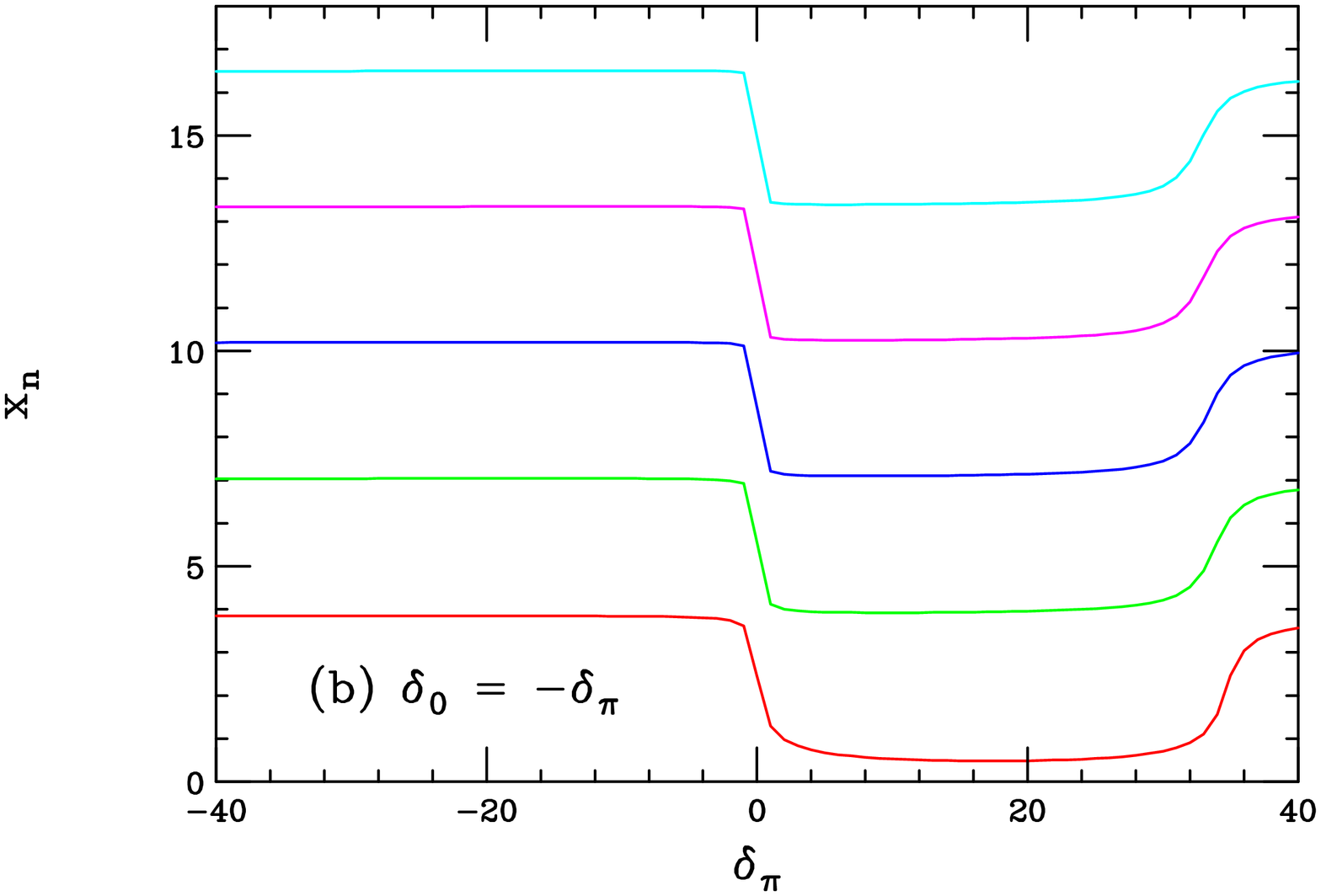}}
\vspace*{0.1cm}
\caption{(a) The behavior of the first five 
roots as a function of $\delta_\pi$ for the case 
$\delta_0=\delta_\pi$. Here, the new root appears at the value 
$\delta_\pi=-kr_c\pi/2\simeq -17.7$.  (b) Same as above for
the case $\delta_0=-\delta_\pi$. }
\label{fig2}
\end{figure}

When $\delta_0\neq 0$ the modifications to the roots are sensitive
to the sign of $\delta_0$.  For $\delta_0=\delta_\pi$, and
in the range $\delta_\pi\gsim - 30$, the roots appear to be very
similar to those obtained when $\delta_0=0$, as is shown in
Fig. \ref{fig2}a.
The major difference in this case is
the location of where the additional ghost root materializes; 
as shown in the Appendix, this root appears at the value 
$\delta_\pi=-(kr_c\pi+\delta_0)$. When $\delta_0>(<)0$ 
the turn-on of this root thus moves to smaller (larger) values
of $\delta_\pi$.  In the case $\delta_0=\delta_\pi$, Fig. \ref{fig2}a
shows that the new root appears at $\delta_\pi=-kr_c\pi/2$ as expected.
We also see that for $\delta_\pi\lsim -\pi kr_c$ the
roots approach those which govern the KK graviton spectrum, as 
discussed
above.   In Fig. \ref{fig2}b, we display the
roots for the case $\delta_0=-\delta_\pi$. For most values of 
$\delta_\pi$
shown in the figure,
these are found to be similar to the case where $\delta_0=0$ except 
that
a new light root does not appear in this case.  This is as expected, 
based
on the discussion in the Appendix.  When $\delta_\pi$ is large and 
positive,
$\delta_\pi\gsim\pi kr_c$, the roots again approach those for the case
of KK gravitons, as anticipated.

\begin{figure}[htbp]
\centerline{
\includegraphics[width=9cm,angle=90]{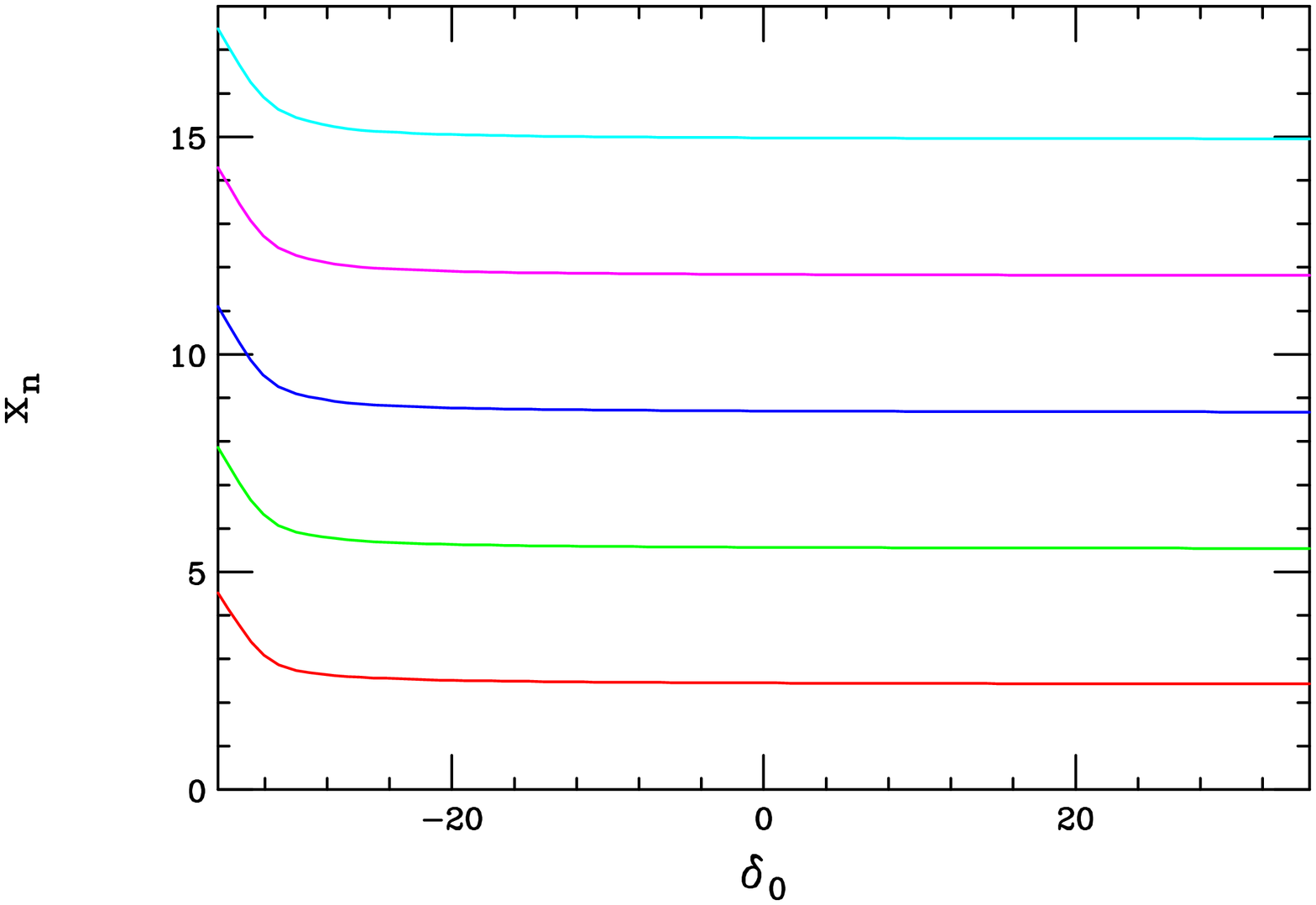}}
\vspace*{0.1cm}
\caption{The behavior of the first five roots as a function of 
$\delta_0$
taking $\delta_\pi=0$.}
\label{fig3}
\end{figure}

When $\delta_\pi=0$, the roots exhibit a weak dependence on
$\delta_0$  away from the $\delta_0=-kr_c\pi$
region; this can be seen in Fig. \ref{fig3}.  
The $\delta_0$ dependence of the roots is isolated 
in the coefficients $\alpha_n$
as seen in Eq.(\ref{alexp}).  From this we see that the roots have a
slightly negative slope with increasing $\delta_0$.


\section{Analysis}

In this section, we study the couplings of the KK modes to
the boundary fermions at $\phi = \pi$.  The values of these 
couplings are the driving force behind the important constraints on
the mass of the lightest KK gauge field in the original RS
scenario \cite{gauge1}.  In this section, we reexamine these bounds
in the presence of brane localized kinetic terms.  To begin, we note 
that the
diagonalization conditions (\ref{diag1}) and (\ref{diag2})
yield the 4-d action
\begin{equation}
S_A = - \frac{1}{4}\int d^4 x
\sum_n \, Z_n \left(F^{\mu \nu(n)}F_{\mu \nu}^{(n)}
- 2 m_n^2 A^{\mu (n)}A_\mu^{(n)}\right)\,,
\label{SZn}
\end{equation}
where we have
\begin{equation}
Z_n = 1 + c_0 [\chi^{(n)}(0)]^2 + c_\pi [\chi^{(n)}(\pi)]^2\,.
\label{Zn}
\end{equation}
Note that for physical fields, we demand that $Z_{n} > 0$ for all
$n\geq 0$.

The coupling of a bulk gauge field $A_\mu$ to fermions $\psi$
localized at $\phi = \pi$ is given by the action
\begin{equation}
S_\psi = - \int d \phi \, d^4 x \,
(det V) \, g_5 \, \bar{\psi} V^\mu_\alpha \gamma^\alpha
A_\mu(x, \phi) \psi \, \delta(\phi - \pi)\,,
\label{Spsi}
\end{equation}
where $V^\mu_\alpha = e^\sigma \eta^\mu_\alpha$ is the
vielbein, $det V = e^{-4\sigma}$, and $g_5$ is the 5-d
coupling constant.  Using the expansion (\ref{AKK}) in
Eq.(\ref{Spsi}), we obtain
\begin{equation}
S_\psi = - \int d^4 x
\left[\frac{g_5}{\sqrt {2 \pi r_c}} \, \bar{\psi} \gamma^\mu
A_\mu^{(0)} \psi + \sum_{n \neq 0} g_5 {e^{\pi kr_c}\zeta_1(x_n)\over
N_n\sqrt{r_c}}\, \bar{\psi} \gamma^\mu
A_\mu^{(n)} \psi \right]\,,
\label{Spsi2}
\end{equation}
where we have performed the redefinition
$\psi \to e^{3 k r_c \pi/2} \psi$ to make the $\psi$ kinetic terms
canonical.  Here, we have used the normalization
\begin{equation}
N_n^2 = {x_n^2\over kr_c\epsilon_n^2}\, \zeta_1^2(x_n) W\,,
\label{Nn}
\end{equation}
where
\begin{equation}
W=(1-2\delta_\pi+\delta_\pi^2x_n^2)\left[ 1 - {\epsilon_n^2
\zeta_1^2(\epsilon_n)[1+2\delta_0+\epsilon_n^2\delta_0^2]\over
x_n^2\zeta_1^2(x_n)[1-2\delta_\pi+x_n^2\delta_\pi^2]}\right]\,.
\label{bigW}
\end{equation}
Note that this normalization is mode dependent.
To bring the gauge field kinetic terms in
Eq.(\ref{SZn}) into the canonical form, we require
\begin{equation}
A^{(n)}_\mu \to \frac{A^{(n)}_\mu}{\sqrt{Z_n}}\,.
\label{Ared}
\end{equation}
Eqs.(\ref{Spsi2}) and (\ref{Ared}) then yield
\begin{equation}
S_\psi = - \int d^4 x
\left[g_0 \, \bar{\psi} \gamma^\mu
A_\mu^{(0)} \psi + \sum_{n \neq 0} g_n
\bar{\psi} \gamma^\mu
A_\mu^{(n)} \psi \right]\,,
\label{Spsi3}
\end{equation}
with 
\begin{equation}
g_0 \equiv {g_5\over \sqrt{2 \pi r_c Z_0}}
\end{equation}
and
\begin{equation}
g_n \equiv g_0\sqrt{2\pi kr_c} \sqrt{{Z_0\over WZ_n}}\,.
\end{equation}  
Here, $g_0$ is identified as the 
usual 4-d coupling constant for the interactions of the
fermions with the zero-mode KK state.  The KK excitation couplings
are given by $g_n$, which due to the presence of the
boundary terms are now mode dependent.  Without the presence of
the brane kinetic terms, one obtains $Z_0=Z_n=1$ which yields the
result that $g_n$ is an approximately mode-independent constant given
by $g_n\simeq g_0\sqrt{2\pi kr_c}$, since in this case 
$\alpha_n\sim 10^{-2}$ and $W\simeq 1$.

\begin{figure}[htbp]
\centerline{
\includegraphics[width=6.5cm,angle=90]{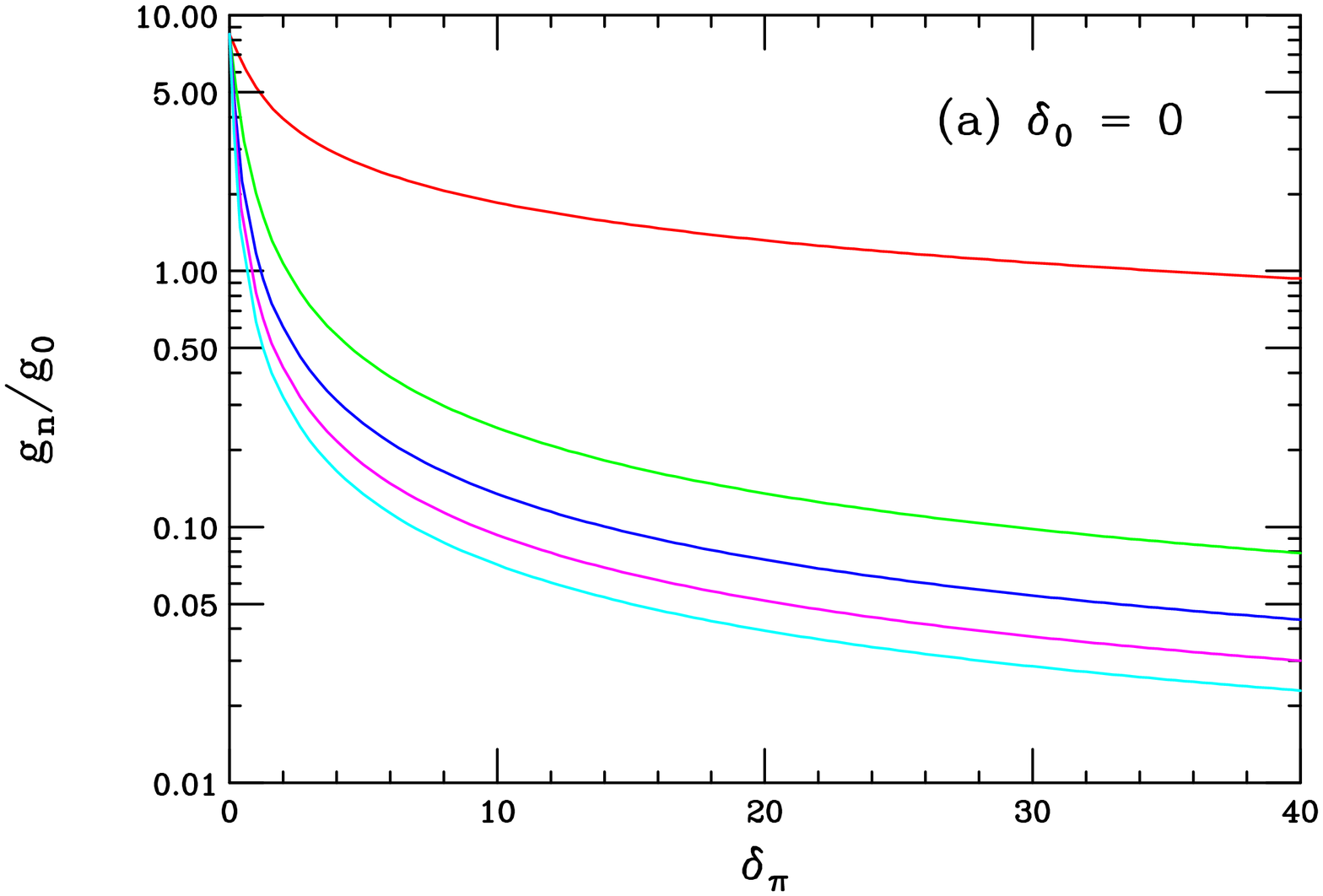}}
\vspace*{5mm}
\centerline{
\includegraphics[width=6.5cm,angle=90]{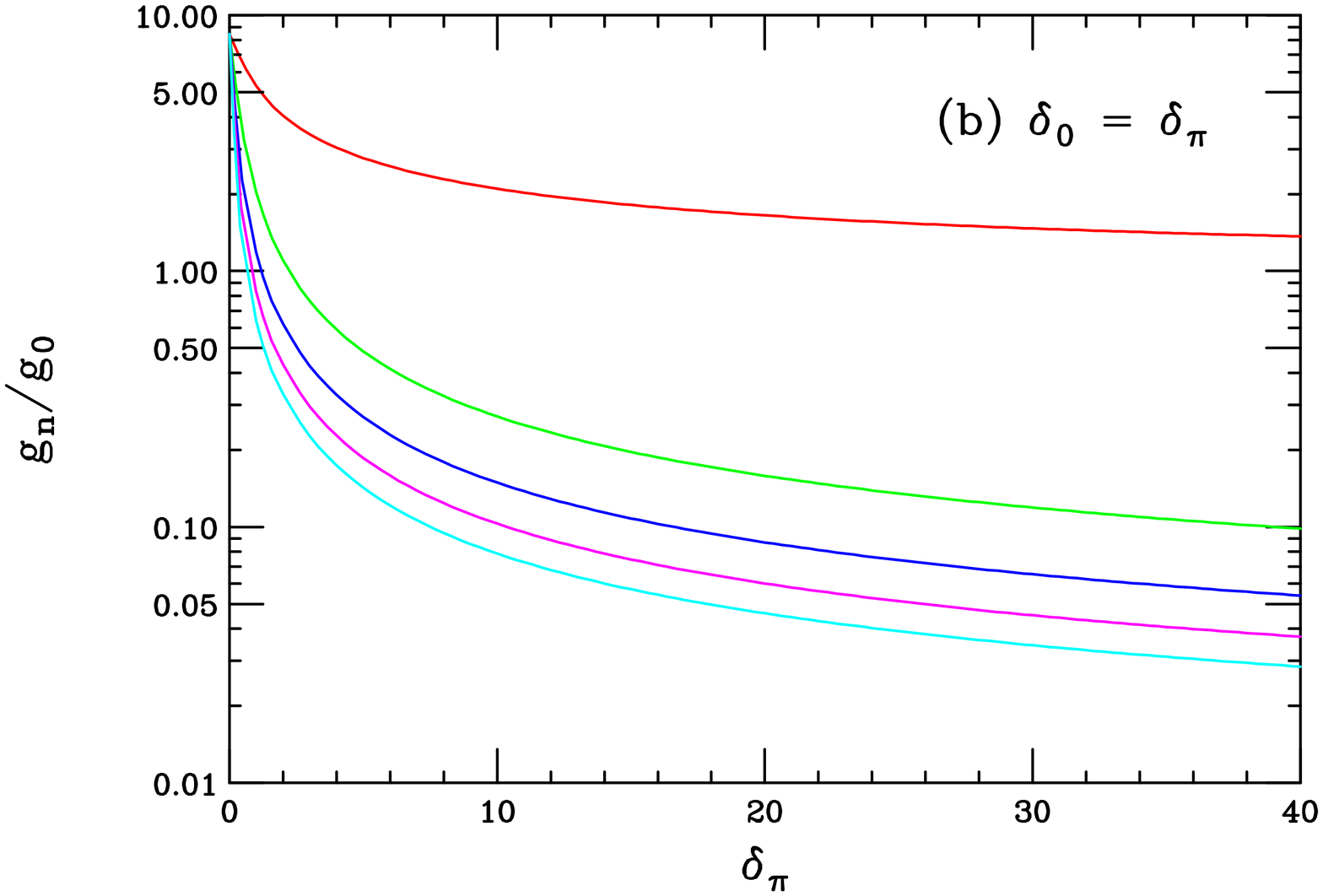}}
\vspace{5mm}
\centerline{
\includegraphics[width=6.5cm,angle=90]{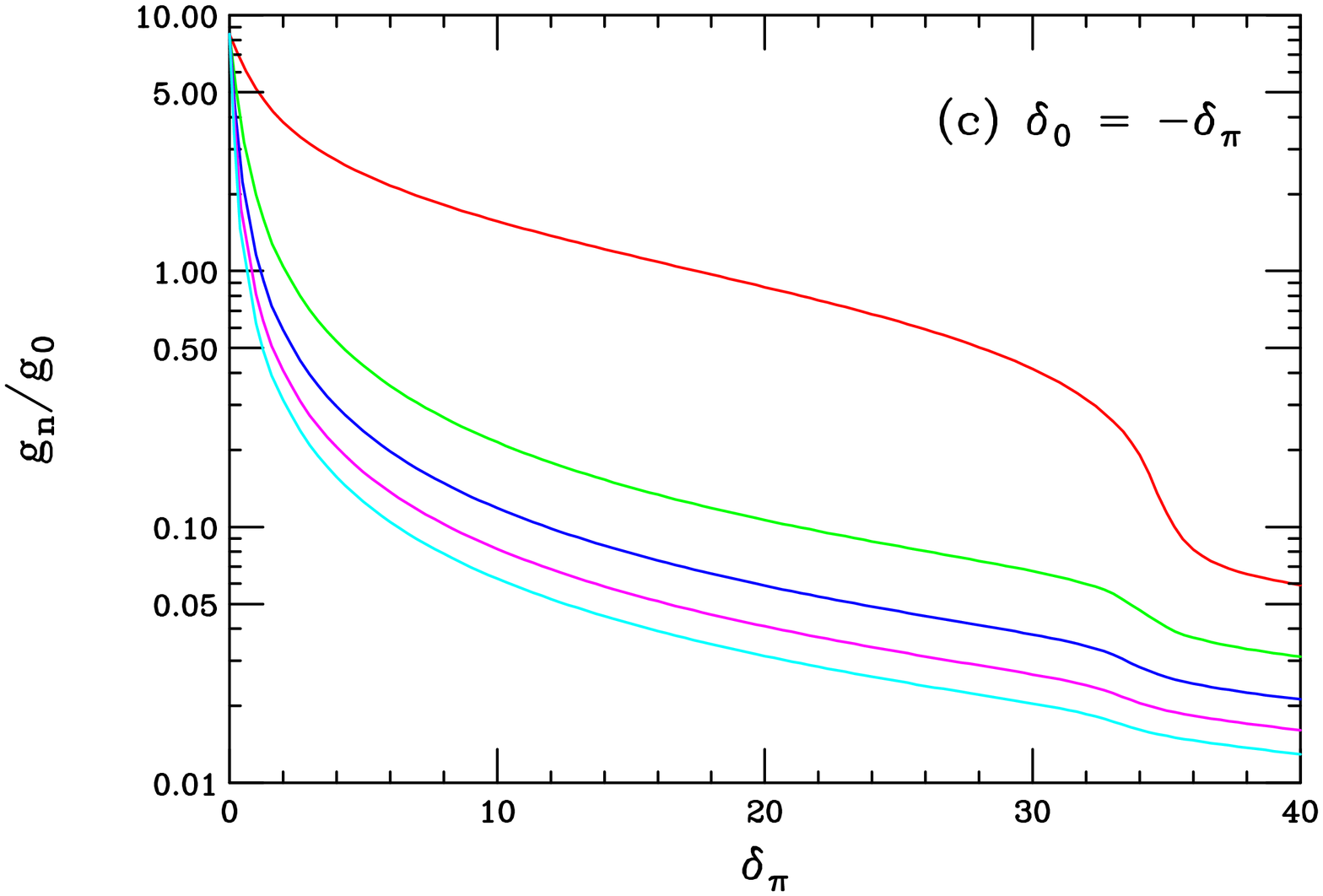}}
\vspace*{0.1cm}
\caption{The ratios of the first five KK couplings to that of the 
zero-mode as
a function of $\delta_\pi$ for the cases
(a) $\delta_0=0$, (b) $\delta_0=\delta_\pi$, and
(c) $\delta_0=-\delta_\pi$.  The lightest (heaviest) KK mode
corresponds to the top (bottom) curve in each case.}
\label{fig4}
\end{figure}

We can now compute the size of the effect of the brane localized 
kinetic
terms on the ratio of couplings $g_n/g_0$.
Using Eq.(\ref{Zn}), we obtain
\begin{equation}
Z_0 = 1 + \frac{\delta_0 + \delta_\pi}{\pi kr_c}\,,
\label{Z0}
\end{equation}
and, neglecting terms of order $\epsilon_n$,
\begin{equation}
Z_n \simeq 1 + W^{-1}\left[ 2\delta_\pi+
{8\delta_0\delta_\pi^2\alpha_n^2\over\pi^2\zeta_0^2(x_n)}\right]\,.
\label{estZn}
\end{equation}
Note that similar to the case of TeV$^{-1}$-sized extra dimensions,
 $Z_n$ is found to be $n$-dependent.  From the requirement that
$Z_0>0$, we obtain the constraint
$\delta_0+\delta_\pi >- \pi kr_c \simeq -35.4$.  If this condition 
isn't
satisfied, the state then has negative norm and is unphysical.

\begin{figure}[htbp]
\centerline{
\includegraphics[width=6.5cm,angle=90]{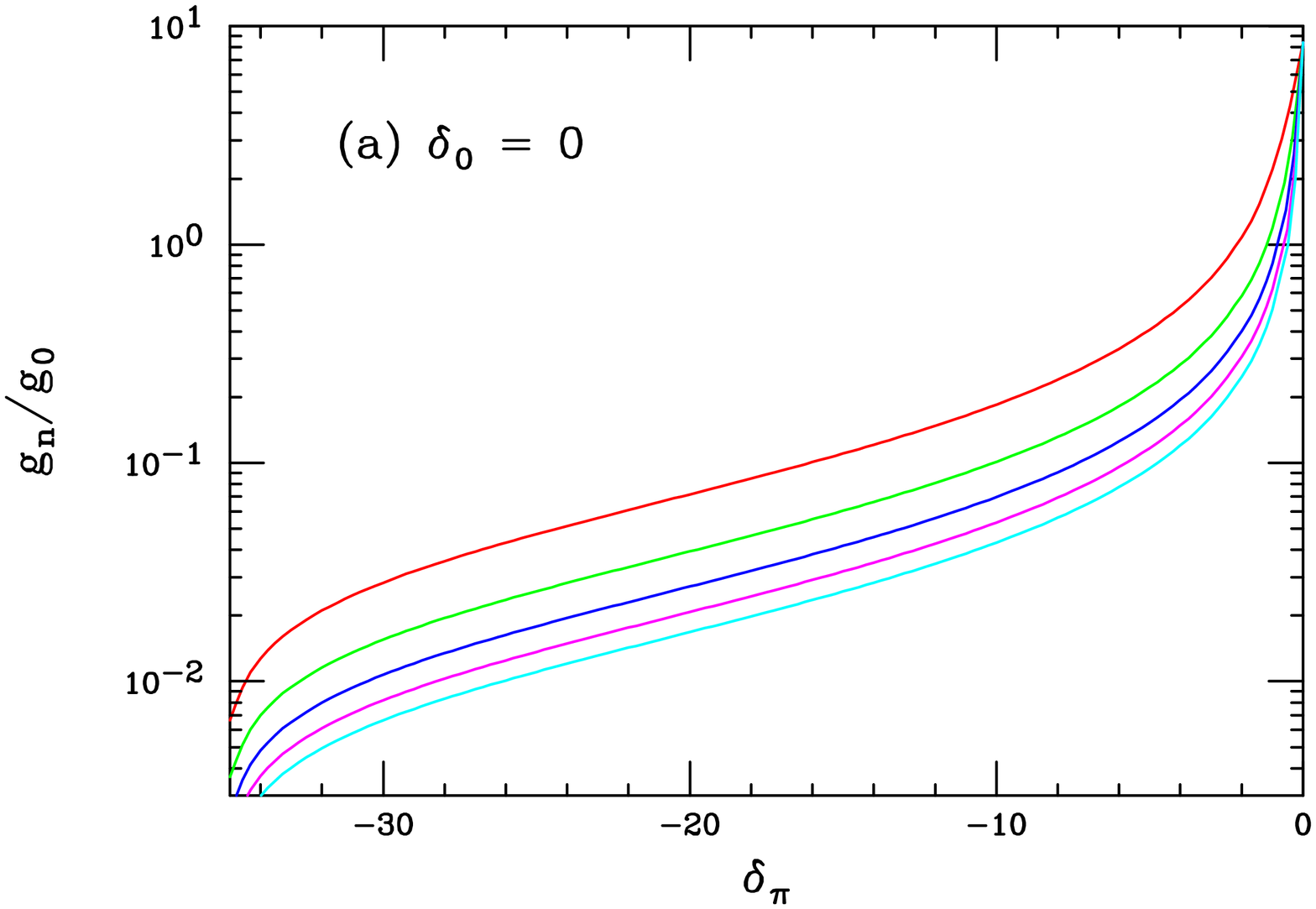}}
\vspace*{5mm}
\centerline{
\includegraphics[width=6.5cm,angle=90]{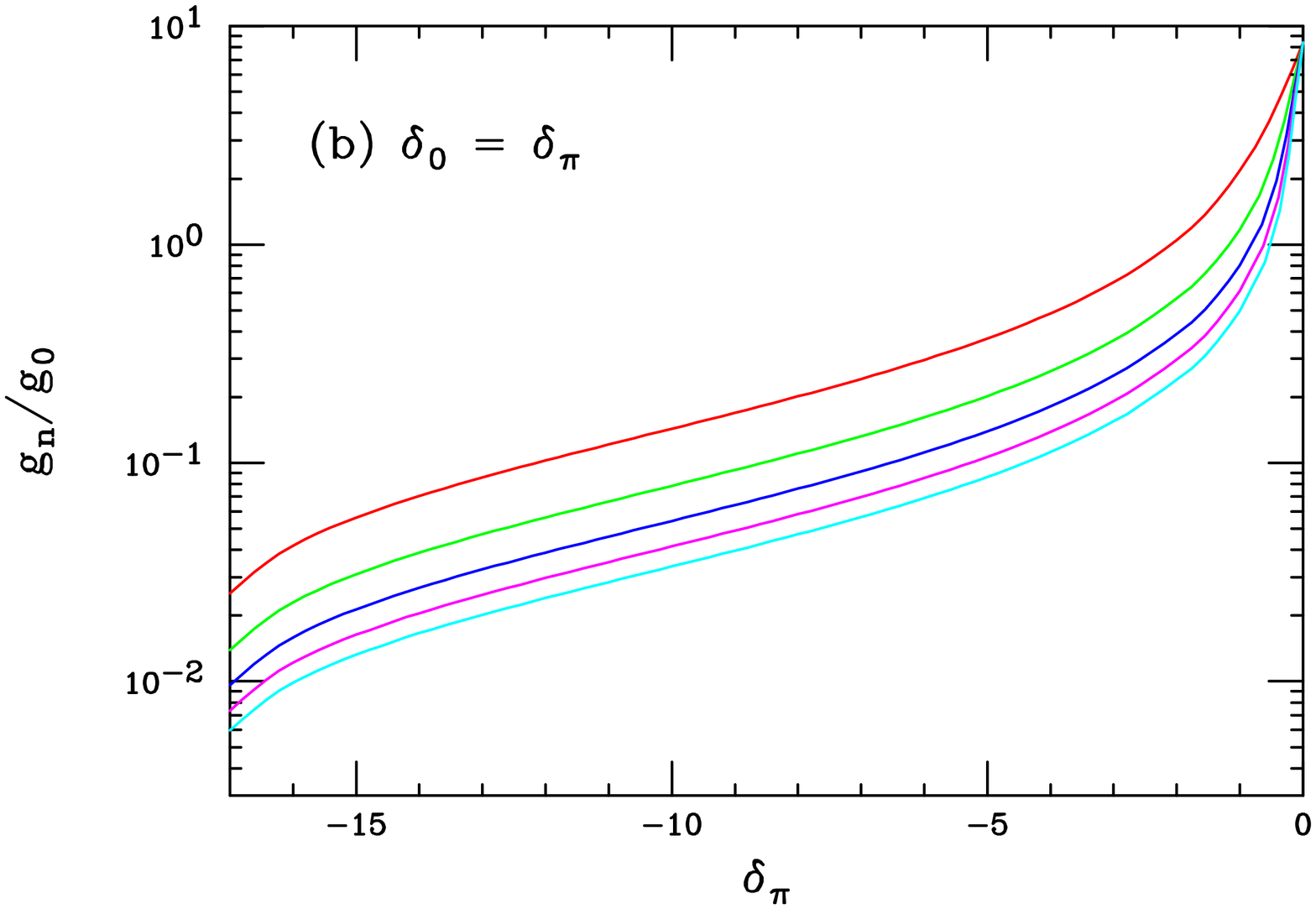}}
\vspace{5mm}
\centerline{
\includegraphics[width=6.5cm,angle=90]{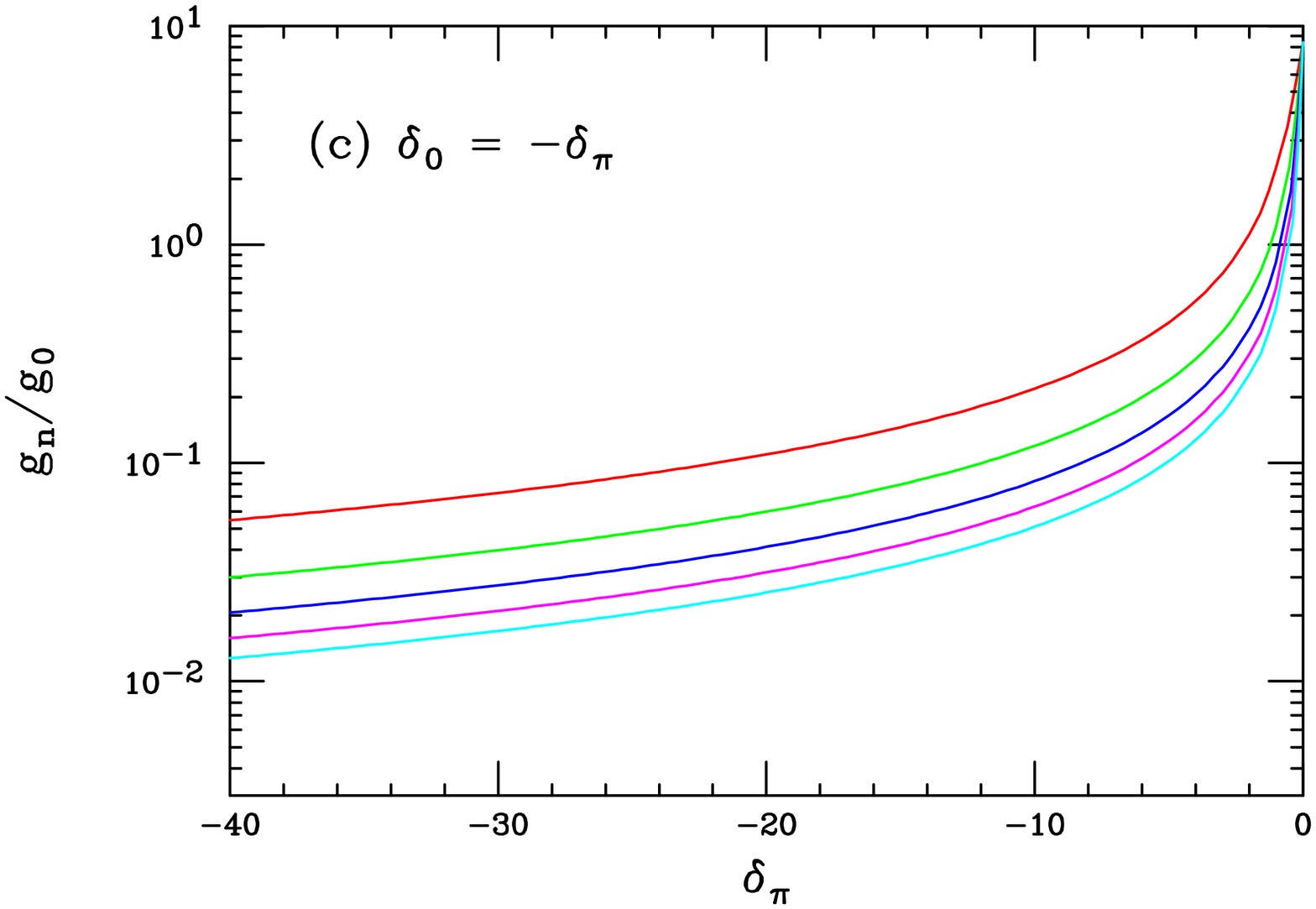}}
\vspace*{0.1cm}
\caption{The ratios of the first five KK couplings to that of the 
zero-mode as
a function of negative $\delta_\pi$ for the cases
(a) $\delta_0=0$, (b) $\delta_0=\delta_\pi$, and
(c) $\delta_0=-\delta_\pi$.  The lightest (heaviest) KK mode
corresponds to the top (bottom) curve in each case.}
\label{newfig}
\end{figure}

We first consider the case  $\delta_\pi\ge 0$, in order
to avoid the ghost states.
Figures \ref{fig4}a-c display the existence of a
significant fall-off of the ratio of couplings
$g_n/g_0$ as a function of $\delta_\pi$.  The values of the coupling 
ratio in the negative $\delta_\pi$ region, away from the regions where
the ghost states appear, are shown in Fig. \ref{newfig}a-c.  In both
cases we see the mode dependence of the coupling strength away from
the value $\delta_\pi=0$.  Note that the coupling strength of the first 
KK
excitation is substantially smaller when $\delta_\pi<0$ compared to
the case with positive values of $\delta_\pi$.
It is clear that compared to 
the original RS framework, the ratio $g_n/g_0$ is 
now naturally much reduced.

\begin{figure}[htbp]
\centerline{
\includegraphics[width=9cm,angle=90]{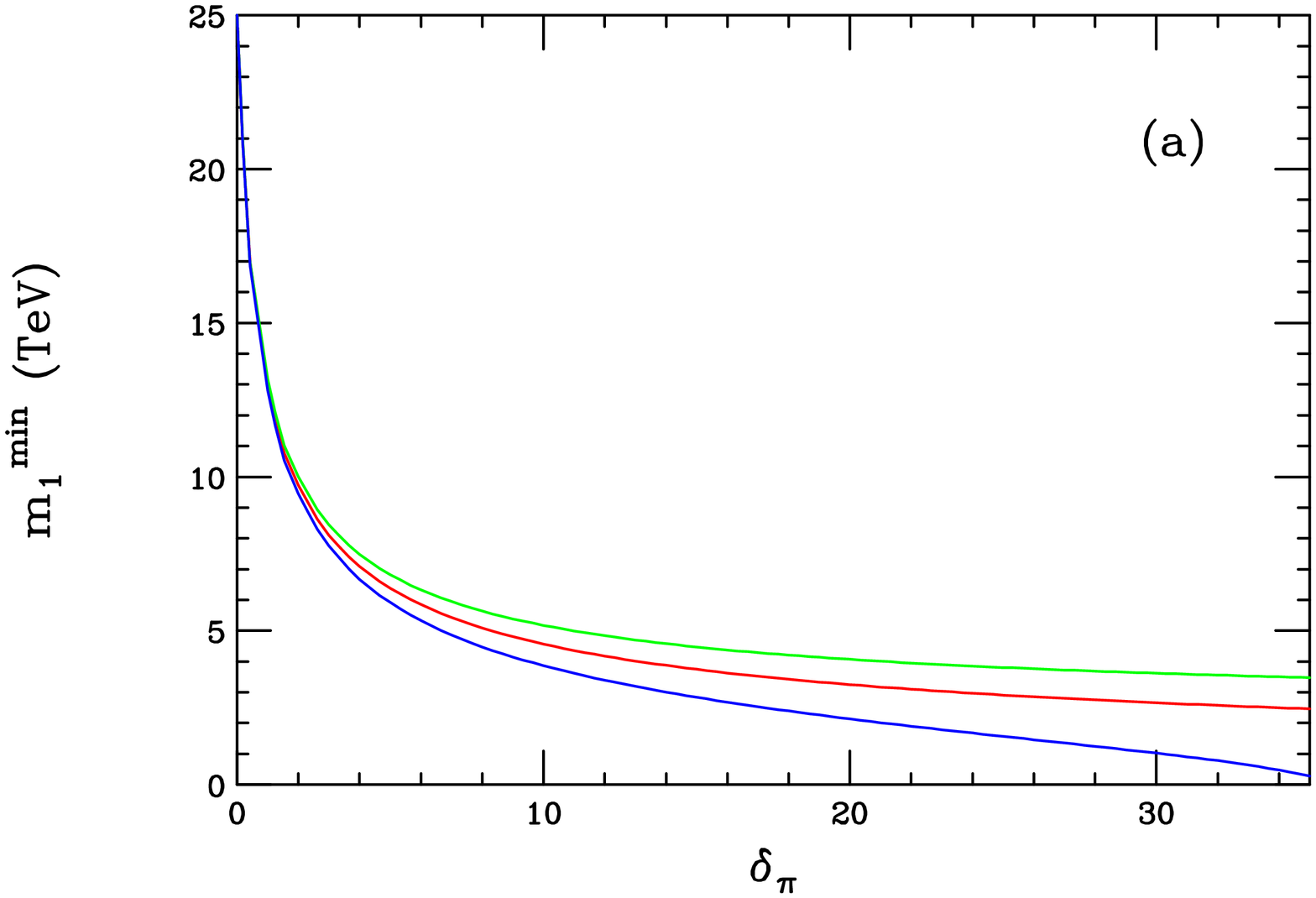}}
\vspace{5mm}
\centerline{
\includegraphics[width=9cm,angle=90]{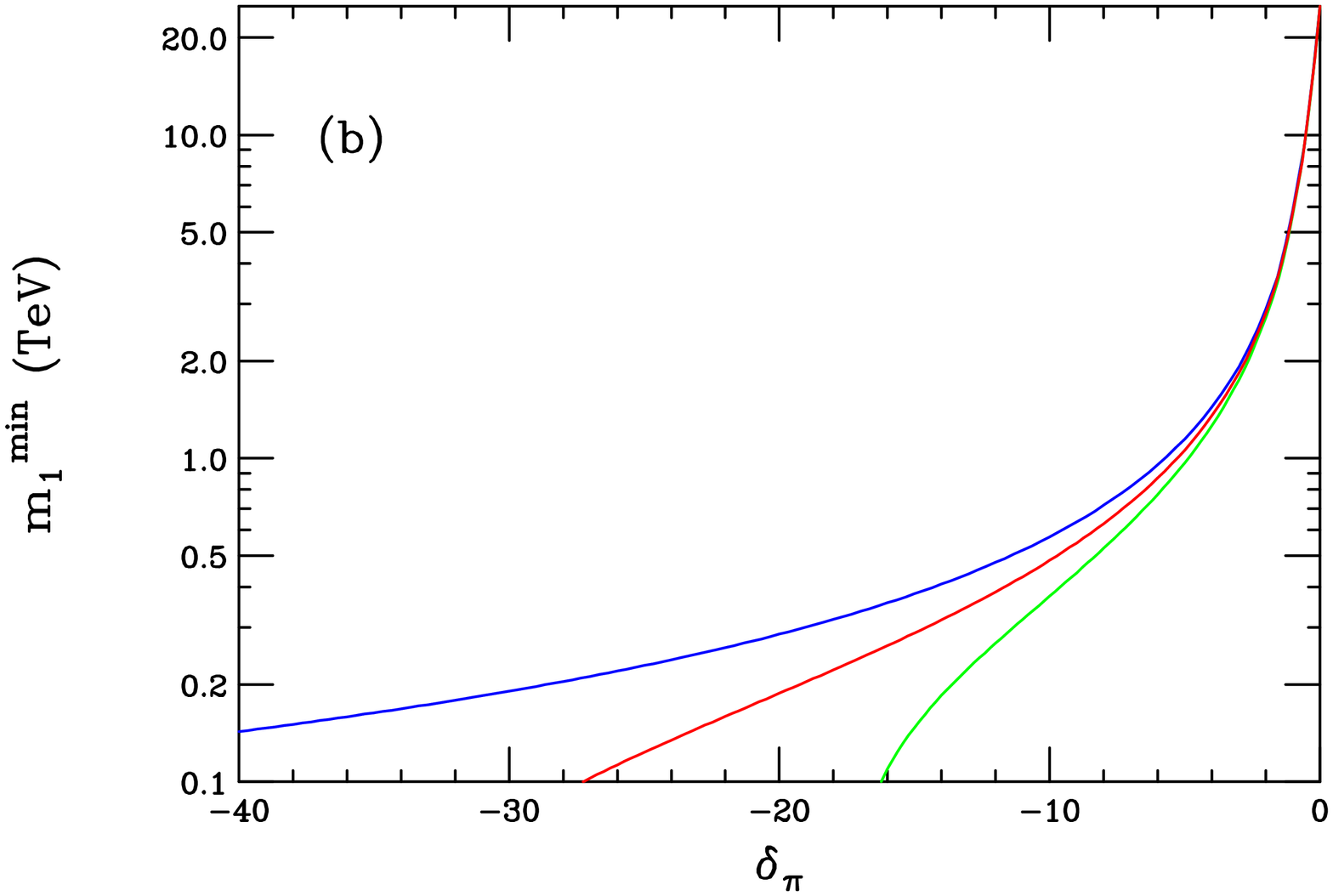}}
\vspace*{0.1cm}
\caption{(a) Lower bound on the mass of the lightest gauge boson KK
excitation as a function of $\delta_\pi\ge 0$ from a fit to the precision
electroweak data for the cases $\delta_0=\delta_\pi$ (top/green),
 $\delta_0=0$ (middle/red), and $\delta_0=
-\delta_\pi$ (bottom/blue).  (b) Same as above but now for the region
$\delta_\pi\le 0$ with $\delta_0=\delta_\pi$ (bottom/green),
 $\delta_0=0$ (middle/red), and $\delta_0=
-\delta_\pi$ (top/blue).}
\label{fig6}
\end{figure}

\begin{figure}[htbp]
\centerline{
\includegraphics[width=9cm,angle=90]{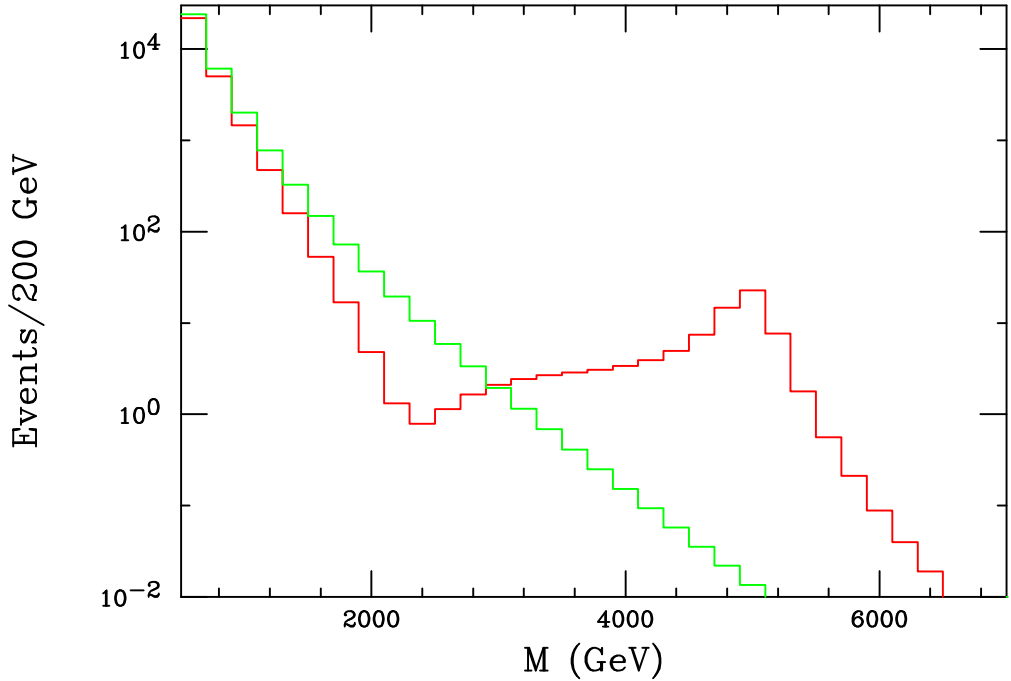}}
\vspace*{0.1cm}
\caption{Drell-Yan event rate at the LHC for the first gauge KK state 
in the
RS scenario with boundary terms. We have assumed a luminosity of
300 \infb\ and have set $m_1=5$ TeV. The steeply falling histogram
represents the SM Drell-Yan continuum.}
\label{fig7}
\end{figure}

Due to the shifts in the mass spectrum and the reduced values of
the ratio $g_n/g_0$,
we now expect that the inclusion of the localized kinetic terms can
result in a significant loosening of the precision electroweak
bounds \cite{gauge1} on the RS model.
To obtain the constraints on the KK gauge boson mass spectrum, 
we need to consider the quantity
\begin{equation}
\tilde V(\delta_\pi,\delta_0)\equiv \sum_n \Big( {g_n\over g_0}\Big)^2
\Big( {x_1\over x_n}\Big)^2\,,
\end{equation}
where the $x_n$ are the roots given above, and examine its detailed
dependence on $\delta_{0,\pi}$. This quantity describes the relative
strength of the set of
contact interactions induced by dimension-6 operators arising
from KK gauge exchanges in precision electroweak observables 
\cite{gauge1}; a full description of the analysis procedure and 
the dependence of
the precision electroweak variables on $V$ can be found in
Ref. \cite{tgrjim}.  Here, we perform a global fit to the most
recent precision electroweak data sample \cite{ewdata} with the
use of ZFITTER.
Since for the case $\delta_{0,\pi}=0$ we know from our previous 
analysis that precision measurements  imply the bound
$m_1|_{RS}\gsim 25$ TeV for the lightest gauge KK excitation 
\cite{gauge1},
we can easily obtain the corresponding constraint for non-zero values of
$\delta_{0,\pi}$.  We obtain the bound $m_1|_{BLKT}\geq m_1|_{RS}
[\tilde V(\delta_\pi,
\delta_0)/\tilde V(0,0)]^{1/2}$; the numerical results of this analysis
is presented in Fig. \ref{fig6}a-b.  Here, we see that the bound on 
$m_1$ falls rapidly as the magnitude of $\delta_\pi$ increases.
For example, with $\delta_\pi\sim 10-20$, we see that $m_1$ can be as 
low as $2-3$ TeV which will lead to an observable signal in Drell-Yan 
production for the first gauge KK excitation
at the LHC. This is demonstrated in Fig \ref{fig7} for the case of
$m_1=5$ TeV.  For larger values of $\delta_\pi$, such states may be
observable at the Tevatron. 
For $\delta_\pi<0$ away from the ghost root region, the
constraints from electroweak data are even weaker; here, direct 
searches at the Tevatron can place bounds on the parameter space.

\begin{figure}[htbp]
\centerline{
\includegraphics[width=9cm,angle=90]{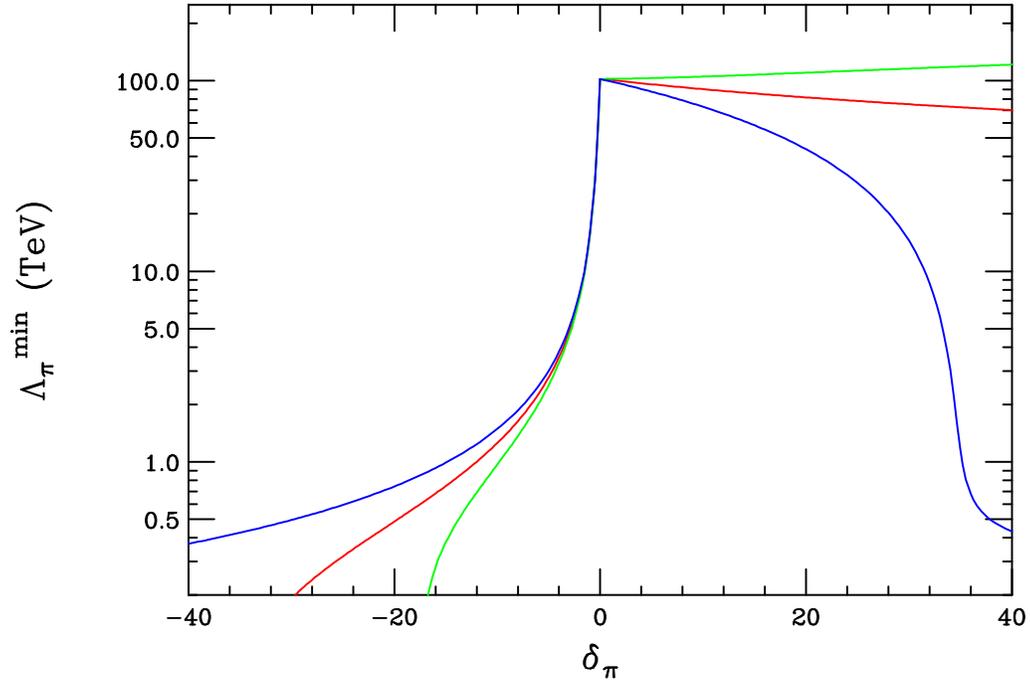}}
\vspace*{0.1cm}
\caption{Bounds on $\Lambda_\pi$ as a function of $\delta_\pi$
for the cases $\delta_0=\delta_\pi$ (green),
 $\delta_0=0$ (red), and $\delta_0=
-\delta_\pi$ (blue), as 
derived from the constraints on the mass of the first gauge KK
state.  The color coding of the curves corresponds to $\delta_0=
\delta_\pi\,,0\,,-\delta_\pi$ being represented by the top(bottom),
middle(middle), bottom(top) curves on the right(left) handside of
the figure.}
\label{fig8}
\end{figure}

So far, we have shown that the inclusion of boundary kinetic terms allow 
for a substantially lighter mass spectrum for the gauge KK states than in
the conventional RS scenario.  However, the question remains as to
whether this ameilorates the stringent bound of
$\Lambda_\pi\gsim 100$ GeV.  To address this issue, we 
convert the constraint on $m_1$ into one on $\Lambda_\pi$ by
noting that $m_1=x_1\Lambda_\pi k/\mpl$, recalling that
$k/\mpl\le 0.1$, and using the values of $x_1$ that we computed above.
Solving for $\Lambda_\pi$, we find the results displayed in 
Fig. \ref{fig8}.  Note that for most values of the parameters, the resulting 
bound on $\Lambda_\pi$ remains quite large for positive values of 
$\delta_\pi$.  This is due to the interplay of the modified roots and
the constraints on the gauge KK mass spectrum.
However, for $\delta_\pi<0$, we see that the lower bound on
$\Lambda_\pi$ is easily reduced by almost an order of magnitude
or more.  This would seem to be the preferred region of parameter
space for this model in its relation to the hierarchy problem.


\section{Conclusions}

A minimal, yet interesting, extension of the RS model is
obtained by assuming that only gauge fields propagate in
the bulk.  However, this setup has been shown to be
constrained by precision electroweak data to have a scale
$\Lambda_\pi \grtsim 100$ TeV for physics on the TeV
brane \cite{gauge1}.  This makes such a scenario less attractive
as a means for resolving the gauge hierarchy.  The stringent bound on 
the
electroweak scale $\Lambda_\pi$ 
is a result of the strong coupling of the gauge
KK modes to the boundary fermions.  It is thus interesting
to examine if a modification of the boundary physics can lead
to a relaxation of this constraint.  One such possibility which is
a simple extension of the
theory, motivated by field theoretic considerations
\cite{ggh}, is the addition of brane localized kinetic terms for the 
bulk
gauge fields.

In this paper, we assumed the presence of localized gauge field
kinetic terms on both the Planck and TeV branes  in the RS model.
We then derived the wavefunctions and the mass spectrum of the gauge KK
modes.   The couplings of the KK states to the SM brane
localized fermions were then calculated.  It was shown that for
natural choices of parameters, a substantial suppression of
these KK couplings, compared to the original model, can
be achieved.  We then reexamined the precision electroweak bounds
on the lightest gauge KK mass or, equivalently, $\Lambda_\pi$.
We found that for a reasonably wide range of parameters, the mass of 
the
lightest gauge KK field can be as low as a few hundred GeV and hence 
will be
visible at the LHC and possibly the Tevatron. 
The weakening of the constraints from precision data
results in a significantly less
severe bound on $\Lambda_\pi$; one can now naturally obtain
$\Lambda_\pi \lsim 10$ TeV.  This makes the RS framework with gauge
fields in the bulk more
favorable, in the context of a solution to the hierarchy
problem, and facilitates the construction of more
realistic models based on the RS proposal.  In addition, for 
much of the $\delta_{0,\pi}$
parameter space, the gauge field and graviton KK excitations 
are nearly degenerate which may result in some interesting collider
phenomenology.

Given the observations in this paper, we expect that the addition of
brane localized kinetic terms 
for other bulk fields, such as gravitons, could significantly alter the
phenomenology of the RS model, and could also lead to the appearance of
new features in the low energy theory.

\bigskip
\noindent{\Large\bf Acknowledgements}

The work of H.D. was supported by the US Department of Energy under
contract DE-FG02-90ER40542.

Note Added: A related article appeared at a similar time \cite{fnal}; 
we thank
these authors for pointing out a sign error in an earlier version of
this manuscript.

\bigskip
\noindent{\Large\bf Appendix}

Here, we will show that near special values of the BLKT
coefficients $\delta_0$ and $\delta_\pi$ 
a vanishing root develops in  Eq.({\ref{xn2}), 
signaling the appearance of an unphysical ghost
state.  We will also discuss the large $|\delta_\pi|$ limit of
this equation. In the limit where $x_n \to 0$, Eqs.(\ref{xn2}) and
(\ref{alexp}) yield
\begin{equation}
x_n^2 = \frac{-4(\delta_\pi + \delta_0 + k r_c \pi)} {-(k r_c \pi
+ \delta_0)(2 \delta_\pi + 1) + (1 + \delta_\pi)}. \label{xnsq}
\end{equation}
This equation is valid for $x_n \ll 1$, and for $\delta_\pi =
-(\delta_0 + k r_c \pi)$, it yields $x_n^2 = 0$.  Thus, for $\delta_\pi
\leq -(\delta_0 + k r_c \pi)$, an unphysical ghost state, a state
with negative norm, appears in
the spectrum, as implied by Eq. (\ref{Z0}).  Note that in the
original RS model, with $\delta_0 = \delta_\pi = 0$, this ghost
state does not appear, since $k r_c \pi > 0$.

We now study Eq.(\ref{xnsq}) in the limit $\delta_\pi \to \pm
\infty$, for the three cases $\delta_0 = 0, \delta_\pi,
-\delta_\pi$ that were examined in the text.  For $\delta_0 = 0$,
we have $x_n^2 \to 4/(2 k r_c \pi - 1)$ as $\delta_\pi \to \pm
\infty$.  Since $k r_c \pi \approx 35$, the small $x_n$
approximation is valid and for $\delta_\pi \to +\infty$, the root
is physical, as shown in Fig. 1a.  Here, we see that the root for
the lowest KK state asymptotes to a constant value of $\simeq 0.24$ as
$\delta_\pi$ becomes large and positive; when $\delta_\pi$ is large
and negative, the ghost root asymptotes to this same value.
For $\delta_0 = \delta_\pi$,
$x_n^2 \to 4/\delta_\pi$ as $\delta_\pi \to \pm \infty$.  This
root is only physical for $\delta_\pi \to +\infty$, and approaches
zero, as presented in Fig. \ref{fig2}a.  In the case where $\delta_0 =
-\delta_\pi$, we have $x_n^2 \to -(4 k r_c \pi)/(2 \delta_\pi^2)$
as $\delta_\pi \to \pm \infty$, and there are no light physical
modes in this limit, as seen in Fig. 2b.

We would like to note that in the limit $\delta_\pi \to \pm
\infty$, with choices of $\delta_0$ that yield $|\alpha_n| \ll 1$,
Eq.(\ref{xn2}) is well-approximated by $J_1(x_n) = 0$, for $x_n
\neq 0$. In this case, apart from the special light modes discussed 
above,
the rest of the gauge field KK modes have masses that are
approximately degenerate with those of the KK gravitons from the
original RS model \cite{phen}. More generally, as can be seen from
Figures \ref{fig1}a and \ref{fig2}, a large range of parameters exist
where this near degeneracy occurs.  This will have important 
phenomenological
implications at future colliders.

\def\MPL #1 #2 #3 {Mod. Phys. Lett. {\bf#1},\ #2 (#3)}
\def\NPB #1 #2 #3 {Nucl. Phys. {\bf#1},\ #2 (#3)}
\def\PLB #1 #2 #3 {Phys. Lett. {\bf#1},\ #2 (#3)}
\def\PR #1 #2 #3 {Phys. Rep. {\bf#1},\ #2 (#3)}
\def\PRD #1 #2 #3 {Phys. Rev. {\bf#1},\ #2 (#3)}
\def\PRL #1 #2 #3 {Phys. Rev. Lett. {\bf#1},\ #2 (#3)}
\def\RMP #1 #2 #3 {Rev. Mod. Phys. {\bf#1},\ #2 (#3)}
\def\NIM #1 #2 #3 {Nuc. Inst. Meth. {\bf#1},\ #2 (#3)}
\def\ZPC #1 #2 #3 {Z. Phys. {\bf#1},\ #2 (#3)}
\def\EJPC #1 #2 #3 {E. Phys. J. {\bf#1},\ #2 (#3)}
\def\IJMP #1 #2 #3 {Int. J. Mod. Phys. {\bf#1},\ #2 (#3)}
\def\JHEP #1 #2 #3 {J. High En. Phys. {\bf#1},\ #2 (#3)}

\end{document}